\documentclass[epj,final]{svjour}
\usepackage{epsfig}

\begin{document}
%\hugehead
\onecolumn
\title{Search for Solar Axions Produced in the $p + d \rightarrow\rm{^3He}+ A$ Reaction}

\author{
A.V.~Derbin$^a$,{\thanks{$^a$derbin@pnpi.spb.ru}} A.S.~Kayunov$^b$ {\thanks{$^b$alekskayunov@pnpi.spb.ru}} and V.
N.~Muratova$^c$ {\thanks{$^c$muratova@pnpi.spb.ru}}}

\date{}
\twocolumn
 \institute { {{St. Petersburg Nuclear Physics Institute, Russian Academy of Sciences,
Gatchina, Russia 188300}}\\
 }

\abstract{A search for the axioelectric absorption of 5.5-MeV solar axions produced in the $p+d\rightarrow \rm{^3He}+\gamma(5.5
\rm{MeV})$ reaction was performed with two BGO detectors placed inside a low-background setup. A model independent limit on
axion-photon and axion-nucleon couplings was obtained:  $|g_{Ae}\times g_{AN}| \leq 3.2\times 10^{-9} (m_A=0)$. Constraints on
the axion-electron coupling constant were obtained for axions with masses in the $(0.1-1.0)$ MeV range: $g_{Ae}\leq
(1.8-9.0)\times 10^{-7}$. The solar positron flux from $A\rightarrow e^-+e^+$ decay was determined for axions with masses $m_A >
2m_e$. Using the existing experimental data on the interplanetary positron flux, a new constraint on the axion-electron coupling
constant for axions with masses in the $(1.2-5.4)$ MeV range was obtained: $g_{Ae} \leq (1-5)\times 10^{-17}$. }

%\DOI: 10.3103/S1062873810060171
\PACS{14.80.Mz,29.40.Mc, 26.65.+t}
\titlerunning{Search for the resonant absorption of solar axions...}
%\authorrunning{A.V.~Derbin, A.I.~Egorov et al.}
\maketitle

\section{INTRODUCTION}
%%%%%%%%%%%%%%%%%%%%%%%%%%%%%%%%%%%%%%%%%%%%%%%%%%%%%%%%%%%%%%%%%%%%%%%%%%%%%%%%%%%%%%%%%%%%%%%

The axion concept was introduced in a theory by Weinberg \cite{Wei78} and Wilczek \cite{Wil78}, who showed that the solution to the problem
of $CP$ conservation in strong interactions, proposed earlier by Peccei and Quinn \cite{Pec77}, should lead to the existence of a neutral
pseudoscalar particle. The original WWPQ axion model contained certain strict predictions for the coupling constants between an axion and
photons ($g_{A\gamma}$), electrons ($g_{Ae}$), and nucleons ($g_{AN}$), and was soon disproved by experiments performed on reactors and
accelerators, and by experiments with artificial radioactive sources \cite{PDG08}.

 Two classes of new theoretical models of an "invisible" axion retained this particle in the form required for solving the $CP$
problem of strong interactions and at the same time suppressed its interaction with matter. These were the models of  "hadronic"
(or KSVZ) axion \cite{Kim79},\cite{Shi80} and DFSZ axion \cite{Zhi80},\cite{Din81} . The scale of Peccei-Quinn symmetry
violation ($f_A$) in both models is arbitrary and can be extended to the Planck mass $m_P \approx 10^{19}$ GeV. The axion mass
in these models is determined by the axion decay constant $f_A$:
\begin{equation}\label{ma}
  m_A{\rm{(eV)}}\approx f_\pi m_\pi /f_A \approx 6.0\times 10^6/{f_A} {\rm{(GeV)}},
\end{equation}
where $m_\pi$ and $f_\pi$ are, respectively, the mass and decay constant of the neutral $\pi$ meson. Since the axion–-hadron and
axion–-lepton interaction amplitudes are proportional to the axion mass, the interaction between an axion and matter is
suppressed.

 The results from present-day experiments are interpreted within these two most popular axion models. The main experimental
efforts are focused on searching for an axion with a mass in the range of $10^{-6}$ to $10^{-2}$ eV. This range is free of astrophysical
and cosmological constraints, and relic axions with such mass are considered to be the most likely candidates for the particles that form
dark matter.

 The effective coupling constants $g_{A\gamma}$, $g_{Ae}$, and $g_{AN}$  are to a great extent  model dependent. For example, the hadronic axion cannot interact directly with leptons, and the
 constant $g_{Ae}$ exists only because of radiative corrections. The constants $g_{A\gamma}$ and  $g_{AN}$ can differ by more than two orders of magnitude from the values accepted in the KSVZ and
 DFSZ models \cite{Kap85}.

New possibilities for solving the $CP$ problem are based on the concept of the existence of a world of mirror particles
\cite{Ber01} and supersymmetry \cite{Hal04}. These models allow for the existence of axions with a mass of about 1 MeV, and this
existence is precluded by neither laboratory experiments nor astrophysical data.

 The purpose of this study is to search experimentally for solar axions with an energy of 5.5 MeV,  produced in
the $p + d \rightarrow\rm{^3He}+ A$  (5.5 MeV) reaction as a result of the axioelectric effect in bismuth atoms. The axion flux
is proportional to the $pp$-neutrino flux, which is known with a high degree of accuracy \cite{Ser09}. The range of axion masses
under study was expanded to 5 MeV.

In previous works we searched for solar axions emitted in the 478 keV M1-transition of $^7{\rm{Li}}$ \cite{Der05}, in the 14.4
keV M1-transition of $^{57}\rm{Fe}$ \cite{Der07,Der09A} and axions produced by the inverse Primakoff conversion of photons in
the electric field of the plasma \cite{Der07A,Der09}.

The results of laboratory searches for the axion as well as the astrophysical and cosmological axion bounds one can find in \cite{PDG08}.

\section{ AXION PRODUCTION IN NUCLEAR
MAGNETIC TRANSITIONS AND THE AXIOELECTRIC EFFECT}
%%%%%%%%%%%%%%%%%%%%%%%%%%%%%%%%%%%%%%%%%%%%%%%%%%%%%%%%%%%%%%%%%%%%%%%%%%%%%%%%%%%%%%%%%%%%%%%
If the axion does exist, the Sun should be an intense source of axions. They could be produced efficiently on the Sun due to photon–-axion
conversion in an electromagnetic field of plasma. Monochromatic axions would be produced in magnetic transitions in nuclei whose low-lying
levels were excited by the high temperature of the Sun.

The reactions of a main solar cycle could also produce axions. The most intensive flux would be expected as a result of the formation of
the $\rm{^3He}$ nucleus:
\begin{equation}
 \rm{p + d\rightarrow{^3He} + \gamma (5.49\;MeV)}.
\end{equation}

According to the standard solar model (SSM), 99.7\% of all deuterium is produced as a result of the fusion of two protons, $p +
p \rightarrow d + e^+ + \nu_e$, while the remaining 0.3\% is due to the  $p+ p + e^- \rightarrow  d + \nu_e$ reaction. The
produced deuteron captures a proton with lifetime $\tau = 6 s$. The expected solar axion flux can thus be expressed in terms of
the $pp$-neutrino flux, which is $6.0\times 10^{10} {\rm{cm}}^{-2} {\rm{s}}^{-1}$ \cite{Ser09}. The proportionality factor
between the axion and neutrino fluxes is determined by the axion-nucleon coupling constant $g_{AN}$, which consists of isoscalar
$g^0_{AN}$ and $g^3_{AN}$ isovector components. The ratio of the probability of a nuclear transition with axion production
$(\omega_A)$ to the probability of a magnetic transition $(\omega_\gamma)$ takes the form \cite{Hax91}-\cite{Avi88}:

\begin{equation}\label{axion_prob}
\frac{\omega_{A}}{\omega_{\gamma}} =
\frac{1}{2\pi\alpha}\frac{1}{1+\delta^2}\left[\frac{g^{0}_{AN}\beta+g^{3}_{AN}}{(\mu_{0}-0.5)\beta+\mu_{3}-\eta}\right]^{2}
\left(\frac{p_{A}}{p_{\gamma}}\right)^{3},
\end{equation}
where $p_{\gamma}$ and $p_{A}$ are, respectively, the photon and axion momenta; $\delta^2 = E/M$ is the probability ratio for the $E$ and
$M$ transitions; $\alpha\approx 1/137$ is the fine-structure constant; $\mu_0 = \mu_p + \mu_n\approx 0.88$ and $\mu_3 = \mu_p - \mu_n
\approx 4.71$ are, respectively, the isoscalar and isovector nuclear magnetic momenta; and $\beta$ and $\eta$ are parameters dependent on
the specific nuclear matrix elements.

 Within the hadronic axion model, the constants $g^0_{AN}$ and $g^3_{AN}$ can be written in terms of axion mass\cite{Kap85},\cite{Sre85}:
\begin{equation}\label{gan0}
g_{AN}^{0}=-4.03\times 10^{-8}(m_A/1 {\rm{eV}}),
\end{equation}
\begin{equation}\label{gan3}
g_{AN}^{3}=-2.75 \times 10^{-8}(m_A/1 {\rm{eV}}).
\end{equation}
The similar relations for $g^0_{AN}$ and $g^3_{AN}$ for the DFSZ axion are model dependent to a greater extent but have the same
order of magnitude. Their numerical values lie in the range of ($0.3 - 1.5$) from the values of these constants for the hadronic
axion.
\begin{figure}
\includegraphics[width=9cm,height=10.5cm]{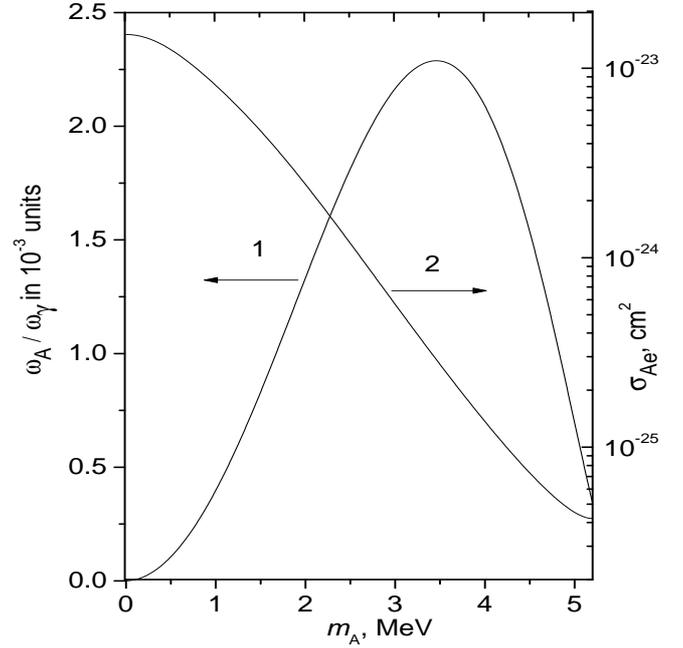}
\caption {Ratio of emission probabilities for axions and $\gamma$ quanta ($\omega_A/\omega_\gamma$) in the $p + d \rightarrow {^3{\rm{He}}}
+ \gamma$ reaction (curve 1, left-hand scale); cross section of the axioelectric effect for 5.5-MeV axions on bismuth atoms for $g_{Ae} =
1$ (curve 2, right-hand scale). } \label{fig1}
\end{figure}

In the $p+d\rightarrow \rm{^3He}+\gamma$ reaction, the M1-type transition corresponds to the capture of a proton with zero
orbital momentum. The probability of proton capture from the $S$ state at proton energies below 80 keV was measured in
\cite{Sch97}; at a proton energy of $\sim 1$ keV, it is $\chi$ = 0.55 $(\delta^2 = 0.82)$. The proton capture from the $S$ state
corresponds to the isovector transition, and the ratio $\omega_A/\omega_\gamma$, which is determined by expression (3),
therefore depends only on $g^3_{AN}$:
\begin{equation}\label{ratio}
\frac{\omega_{A}}{\omega_{\gamma}} =
 \frac{\chi}{2\pi\alpha}\left[\frac{g_{AN}^{3}}{\mu_3}\right]^2\left(\frac{p_A}{p_\gamma}\right)^3 = 0.54(g_{AN}^{3})^2
 \left(\frac{p_A}{p_\gamma}\right)^3.
\end{equation}

The calculated values of the $\omega_A/\omega_\gamma$ ratio as a function of the axion mass are
shown in Fig.\ref{fig1}. The axion flux on the Earth's surface is
\begin{eqnarray}\label{FluxA}
\nonumber\Phi_A = \Phi_{\nu p p}(\omega_A/\omega_\gamma) = 2.7\times
10^{10}(g^3_{AN})^2(p_A/p_\gamma)^3 =\\
= 2.04\times10^{-5}m_A^2(p_A/p_\gamma)^3
\end{eqnarray}
where $\Phi_{\nu p p} = 6.0 \times 10^{10} {\rm{cm}}^{-2} {\rm{s}}^{-1}$ is the $pp$-neutrino flux and $m_A$ is the axion mass
in eV units.

 To detect 5.5-MeV axions, we chose the reaction of axioelectric effect - $A + Z + e
\rightarrow Z + e$. The cross section for bismuth atoms exceeds that of the Compton conversion of an axion ($A +e^-
\rightarrow\gamma + e^-$) by almost two orders of magnitude; the detection efficiency for the produced electron is close to
100\%; and, finally, the background level at 5.5 MeV is much lower than in the range of natural radioactivity. As a result, the
sensitivity to constants $g_{Ae}$ and $g_{AN}$ can be high even in an experiment with a relatively small target mass.

In the axioelectric effect (an analog of the photoelectric effect), an axion disappears and an electron with an energy of $E_e=E_A - E_b$,
where $E_b$ is the electron binding energy, is emitted from the atom. The axioelectric effect cross section for K-shell electrons was
calculated (on the assumption that $E_A\gg E_b$ and $Z\ll 137$) in \cite{Zhi79}:
\begin{center}
\begin{eqnarray}\label{sigmaAE}
\nonumber \sigma_{Ae} = 2(Z\alpha m)^5\frac{g^2_{Ae}}{m^2}\frac{p_e} {p_A}\ [\frac{4E_A(E^2_A+m^2_A)}{(p^2_A- p^2_e)^4}-\\
\nonumber-\frac{2E_A}{(p^2_A -p^2_e)^3} -\frac{64}{3}p^2_ep^2_Am\frac{m^2_A}{(p^2_A -p^2_e)^6}-\frac{16m^2_Ap^2_AE_e}{(p^2_A-p^2_e)^5}-\\
-\frac{E_A}{p_ep_A}\frac{1}{(p^2_A-p^2_e)^2}\ln\frac{p_e+p_A}{p_e-p_A} ].
\end{eqnarray}
\end{center}

The dependence of the cross section on the axion mass for the coupling constant $g_{Ae} = 1$ is shown in Fig. 1. The cross
section depends on the nuclear charge according to the $Z^5$ law, and it is therefore reasonable to search for this process
using detectors with large $Z$. The K-shell electrons make the main contribution to the cross section. The contribution from the
other electrons was incorporated by introducing a factor of $5/4$, by analogy with the photoelectric effect.

\section{ INTERACTION OF AXIONS WITH SOLAR MATTER AND AXION DECAYS}
%%%%%%%%%%%%%%%%%%%%%%%%%%%%%%%%%%%%%%%%%%%%%%%%%%%%%%%%%%%%%%%%%%%%%%%%%%%%%%%%%%%%%%%%%%%%%%%

The flux of 5.5 MeV axions on the Earth-s surface is proportional to the $pp$-neutrino flux only when the axion lifetime exceeds
the time of flight from the Sun and when the flux is not reduced as a result of axion absorption by solar matter. The axions
produced at the center of the Sun must pass through a layer of $\approx 7\times 10^{35}$ electrons per ${\rm{cm}}^{-2}$ in order
to reach the Sun's surface. The Compton conversion of an axion into a photon imposes an upper limit on the sensitivity of
Earth-bound experiments to the constant $g_{Ae}$. The cross section of this reaction for 5.5-MeV axions depends weakly on the
axion mass and can be written as $\sigma_{cc}\approx g^2_{Ae}4\times 10^{-25}{\rm{cm}}^2$. For $g_{Ae}$ values below $10^{-6}$,
the axion flux is not substantially suppressed.

As can be seen in Fig. 1, the maximum cross section of the axioelectric effect on bismuth atoms is $\sigma_{Ae}\approx
g^2_{Ae}1.5\times 10^{-23}{\rm{cm}}^2$. The abundance of heavy ($Z > 50$) elements in the Sun is $\sim10^{-9}$ in relation to
hydrogen. If  $g_{Ae}< 10^{-3}$, the change in the axion flux does not exceed 10\%.
\begin{figure}
\includegraphics[width=9cm,height=10.5cm]{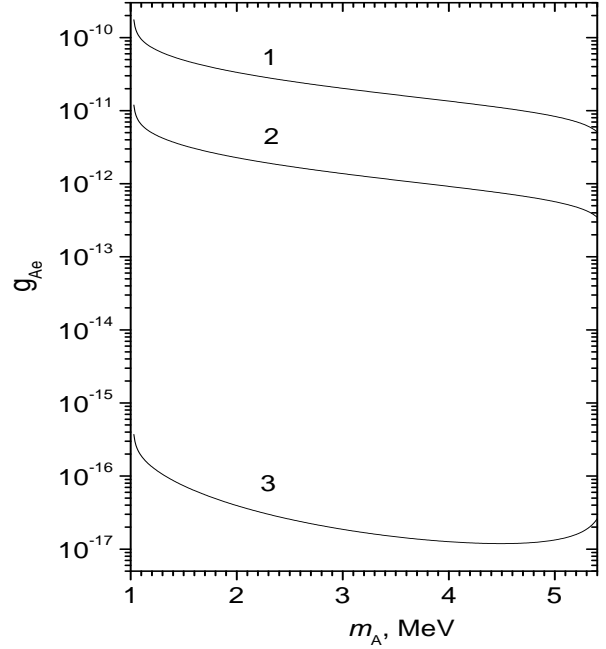}
\caption {Constraints on $g_{Ae}$ and $m_A$ from $A \rightarrow e^+ + e^-$ decay. Curves 1 and 2 were obtained in accordance
with the condition that 90\% of all solar axions reach the Sun's surface and the Earth's surface, respectively. Curve 3
corresponds to the observed positron flux entering the Earth's atmosphere. The allowed $g_{Ae}$ and $m_A$ values are below the
corresponding curves. } \label{fig2}
\end{figure}
The axion-photon interaction, as determined by the constant $g_{A\gamma}$, leads to the conversion of an axion into a photon in
a nuclear field. The cross section of the reaction on a proton is $\sigma_{c.c}\approx g^2_{A\gamma}Z^22\times
10^{-29}{\rm{cm}}^2$, and the condition for the axion emission from the Sun imposes a constraint on $g_{A\gamma}$ : $g_{A\gamma}
< 10^{-4} {\rm{GeV}}^{-1}$. The constraint for the other elements does not grow stricter, due to their low concentration in the
Sun.

The axion-nucleon interaction leads to axion absorption in a reaction similar to photo--dissociation: $A+ Z \rightarrow Z_1 +
Z_2$. This reaction is of a threshold nature; for 5.5-MeV axions, it might occur only for a few nuclei: $^{17}{\rm{O}}$,$
^{13}{\rm{C}}$, and $^2{\rm{H}}$. It was shown in \cite{Raf82} that axio--dissociation cannot substantially reduce the axion
flux for $g_{AN} < 10^{-3}$.

For axions with a mass above $2m_e$, the main decay mode is related to their disintegration into an electron-positron pair: $A
\rightarrow e^+ + e^-$. The lifetime of an axion in the intrinsic reference system has the form:
\begin{equation}
\tau_{cm}=8\pi/(g^2_{Ae}\sqrt{m^2_A - 4m^2_e}).
\end{equation}
The probability of an axion reaching Earth is
\begin{equation}
P(m_A,p_A)=\exp(-\tau_{as}/\tau_{cm}),
\end{equation}
where $\tau_{as}=Lm_A/cp_A$ is the time of flight in the reference system associated with the axion and $L = 1.5\times 10^{13}$
cm is the distance from the Earth to the Sun. The condition $\tau_{as}<0.1\tau_{cm}$ (in this case, $90\%$ of all axions reach
Earth) yields the sensitivity limits for the constant $g_{Ae}$ in our experiment (Fig. 2, curve 2).

The positron flux with an energy of ~5 MeV near the surface of the Earth's atmosphere surface is
$\approx10^{-4}{\rm{cm}}^{-2}{\rm{sr}}^{-1}$. Using this value and expression (5) for $\omega_A/\omega_\gamma$, we can obtain
the upper constraint on the coupling constant $g_{Ae}$ (Fig. 2, curve 3). Curve 1 in Fig. 2 is plotted in accordance with the
condition that 90\% of all solar axions emerge from the Sun. The range of excluded $g_{Ae}$ and $m_A$ values thus lies between
curves 1 and 3. The obtained upper limits  on the constant $g_{Ae}$ $(g_{Ae} < 10^{-16} - 10^{-17})$ for axions with mass in the
$(1.2-5.4)$ MeV range are the strongest up-to date.

If the axion mass is less than $2m_e$, $A\rightarrow e^+ + e^-$ decay is impossible, but the axion can decay into two $\gamma$
quanta. The probability of decay, which depends on the axion–-photon coupling constant and the axion mass, is given by the
expression
\begin{equation}
\tau_{A\rightarrow\gamma\gamma}=64\pi/g^2_{A\gamma}m^3_A.
\end{equation}

The present-day experimental constraint on $g_{A\gamma}$ is $10^{-8}$ ${\rm{GeV}}^{-1}$, which corresponds to $\tau_{cm} = 10^3$
s for 1-MeV axions. This means that the axion flux is not reduced practically due to the $A\rightarrow 2\gamma$ decay for axions
with masses below 1 MeV.

\section{EXPERIMENTAL SETUP}
%%%%%%%%%%%%%%%%%%%%%%%%%%%%%%%%%%%%%%%%%%%%%%%%%%%%%%%%%%%%%%%%%%%%%%%%%%%%%%%%%%%%%%%%%%%%%%%
We used two scintillation detectors (BGO1 and BGO2), manufactured from orthogermanate bismuth ${\rm{Bi}}_3{\rm{Gå}}_4{\rm{O}}_{12}$, to
search for the 5.5 MeV axions. The BGO crystal, which had a mass of 580 g, was shaped as a prism 50 mm in height, with a cross section in
the form of a regular hexagon inscribed in a circle 50 mm in diameter. The detector signal was measured by an PMT-176 photoelectron
multiplier, which had an optical contact with a crystal end surface.

External $\gamma$ activity was suppressed using passive shield that consisted of successive layers of lead (50 mm), iron (35
mm), copper (10 mm), and bismuth (15 mm ${\rm{Bi}}_2{\rm{O}}_3$). The total thickness of the passive shield was $\approx$ 100
${\rm{g\;  cm}}^{-2}$. The setup was located on the Earth's surface. To suppress the cosmic-ray background, an active shielding
composed of five plastic scintillators $500 \times 500 \times 120$ mm in size was used. The total count rate from the active
shielding was set at a level of 600 ${\rm{s}}^{-1}$, which led to $4\%$ dead time at an inhibit pulse width of 70 $\mu s$.

The spectrometric channel of the BGO scintillation detector included an amplifier with a shaping time of 1 $\mu$s and a 12-digit ADC. The
amplification was selected so that the ADC channel scale was 3.25 keV. Standard calibration sources ($^{60}{\rm{Co}}$ and
$^{207}{\rm{Bi}}$), in combination with the natural radioactivity lines of $^{40}{\rm{K}}$ and the uranium and thorium families, were used
for energy calibration of the detector. The energy dependence of detector resolution $\sigma$ can be presented as $\sigma/E \approx 6.3\%
\times E^{-1/2}$, where $E$ is in MeV. The range of 5.5-MeV electrons in the BGO crystal was $\approx$ 3 mm, leading to a loss of electron
detection efficiency near the detector surface. Based on this electron range, our estimate of the detection efficiency was $\varepsilon =
0.87$.
\begin{figure}
\includegraphics[width=9cm,height=10.5cm]{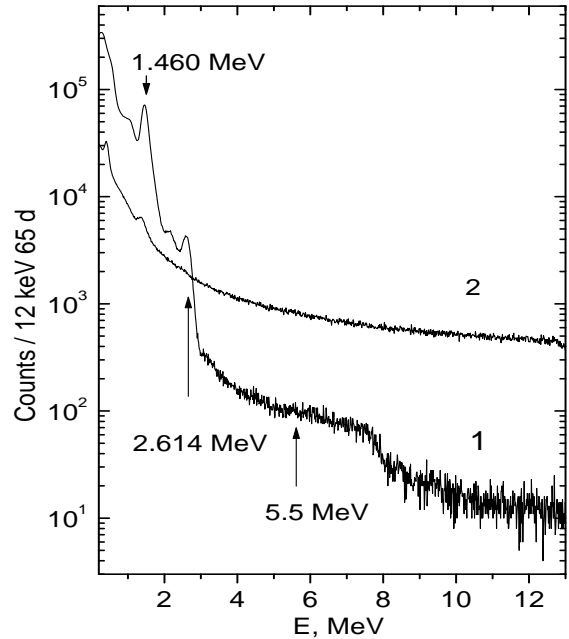}
\caption {The energy spectrum of the BGO1 detector measured (1) in anticoincidence and (2) in coincidence with the active shielding signal.
The arrows indicate peaks with energies of 1.460 MeV, 2.614 MeV, and 5.5 MeV (desired peak). } \label{fig3}
\end{figure}

\section{RESULTS}
%%%%%%%%%%%%%%%%%%%%%%%%%%%%%%%%%%%%%%%%%%%%%%%%%%%%%%%%%%%%%%%%%%%%%%%%%%%%%%%%%%%%%%%%%%%%%%%%%
The measurements were performed over 65 days in real time by series, each of which was 2 h long. The measurements were divided into series
so as to monitor the time stability of the BGO detector and the active shielding. The energy spectrum of the BGO1 detector in the range of
1–-13 MeV is shown in Fig. 3. The spectrum of the BGO1 signals that were not accompanied by the active shielding signal is designated as 1.

In the spectrum, we can identify two pronounced peaks at 1.460 and 2.614 MeV; these are due to the natural radioactivity of the
$^{40}{\rm{K}}$ (located in the PMT's glass housing) and of $^{208}{\rm{Tl}}$ from the $^{232}{\rm{Th}}$ family (Fig. 3).  The positions
and intensities of these peaks were used for monitoring of time stability.

Bismuth has the largest nuclear charge among the stable isotopes (Z = 83), and the cross section of $(e^+e^-)$-pair production upon the
interaction of $\gamma$ quanta is therefore the largest for this element. The annihilation peak at 0.511 MeV is pronounced in the spectrum.
The peak at 2.1 MeV is related to the emission of one annihilation $\gamma$ quantum from the detector upon the detection of 2.614 MeV
$\gamma$ rays. The kink at ~7.5 MeV is due to the $\gamma$ quanta produced as a result of the capture of thermal neutrons by the nuclei of
the iron, copper, and lead components of the passive shield.

The positions and dispersion of the 1.46 and 2.614 MeV peaks determined during the measurements were used to find the energy scale and
energy dependence of the BGO resolution. The energy calibration of the spectrometric channel was found as a linear function: $E = A\times N
+ B$, where $E$ is the released energy and $N$ is the channel number. The dependence of the energy resolution of a scintillation detector
vs energy can be written as $\sigma = C\times \sqrt{E}$. The parameter C was found to be 0.063 ${\rm{MeV}}^{1/2}$ and 0.068
${\rm{MeV}}^{1/2}$ for the first and second detectors, respectively. The values of $\sigma$ determined from the background spectrum are in
good agreement with measurements performed with a $^{207}{\rm{Bi}}$ calibration source. The expected dispersion of the 5.5 MeV peak due to
axion absorption is $\sigma =$ 148 keV and 160 keV for BGO1 and BGO2, respectively.

\begin{figure}
\includegraphics[width=9cm,height=10.5cm]{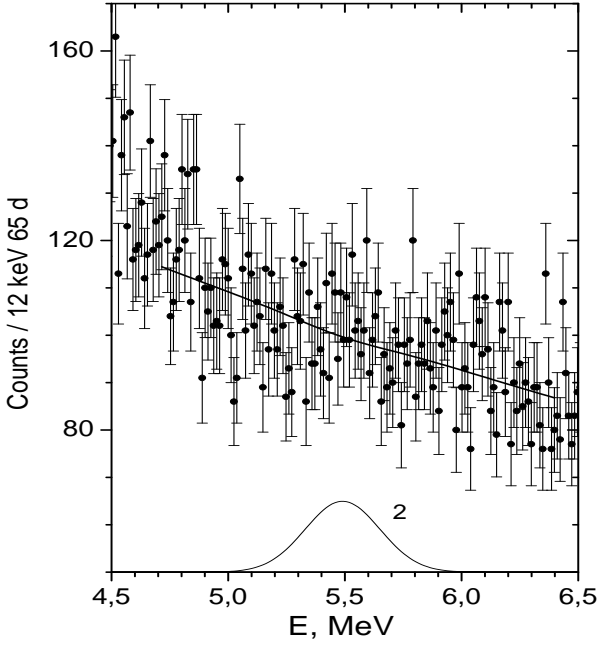}
\caption {Results from fitting the BGO1 spectrum in the range of ($4.5-6.5$) MeV. Curve 2 corresponds to the detector response
function ($E_0 = 5.49$ MeV, $\sigma = 0.148$ MeV). } \label{fig4}
\end{figure}
Figure \ref{fig4}  shows the energy range of ($4.5-6.5$) MeV, in which the axion peak was expected. Since the energy resolutions
of the detectors differ by 10\%, the BGO1 and BGO2 spectra were analyzed separately. The spectrum measured in the range of
$4.7-6.4$ MeV was fitted by a sum of linear and Gaussian functions; the Gaussian peak position and dispersion corresponded to
the desired-peak parameters:
\begin{equation}
N(E) = a+bE+\frac{S}{\sqrt{2\pi\sigma}}exp[-\frac{(E_0-E)^2}{2\sigma^2}] ,
\end{equation}
where $E_0 =$ 5.49 MeV is the axion peak position, $\sigma =$ 0.148 (0.160) MeV is the Gaussian peak dispersion, $S$ is the peak area, and
$a$ and $b$ are the parameters of the function describing the continuous background.

The position peak and dispersion were fixed and three parameters were varied, two of which described the continuous background while the
third described the area of the desired peak. The total number of the degrees of freedom in the range of 4.5–-6.5 MeV was 134 (139). The
fit results for BGO1, corresponding to the minimum $\chi^2 = 139/134$, are shown in Fig. 4. The intensity of the 5.5 MeV peak was found to
be $S = -44 \pm 78$ ($S = -4 \pm 81$ for BGO2). Combining these results, we obtain $S = -48 \pm 111$ for the axion peak area; this
corresponds to the upper limit on the number of counts in the peak, $S_{lim} = 140$ at a 90\% confidence level \cite{Fel98}.

 The expected number of axioelectric absorption events was
\begin{equation}
S_{abs} = \varepsilon N_{Bi}T\Phi_A\sigma_{Ae}
\end{equation}
where $\sigma_{Ae}$ is the axioelectric effect cross section, given by expression (\ref{sigmaAE}); $\Phi_A$ is the axion flux
(\ref{FluxA}); $N_{Bi} = 1.88 \times 10^{24}$ is the number of Bi atoms; $T = 5.62\times10^6$ s is the measurement time; and $\varepsilon =
0.87$ is the detection efficiency for 5.5 MeV electrons. Axion flux $\Phi_A$ is proportional to the constant $g^2_{AN}$ , and the cross
section $\sigma_{Ae}$ is proportional to the constant $g^2_{Ae}$ , according to expressions (\ref{FluxA}) and (\ref{sigmaAE}). As a result,
the $S_{abs}$ value depends on the product of the axion-electron and axion-nucleon coupling constants: $g^2_{Ae}\times g^2_{AN}$.

The experimentally found condition $S_{abs} \leq S_{lim}$ imposes some constraints on the range of possible $g_{Ae}\times
g^3_{AN}$  and $m_A$ values. The range of excluded $|g_{Ae}\times g^3_{AN}|$ values is shown in Fig. 5, at $m_A \rightarrow 0$
the limit is $|g_{Ae}\times g^3_{AN}| \leq 3.2\times10^{-9}$. The dependence of $|g_{Ae}\times g^3_{AN}|$ on $m_A$ is related
only to the kinematic factor in formulas (\ref{ratio}) and (\ref{sigmaAE}). These constraints are completely model-independent
and valid for any pseudoscalar particle.

\begin{figure}
\includegraphics[width=9cm,height=10.5cm]{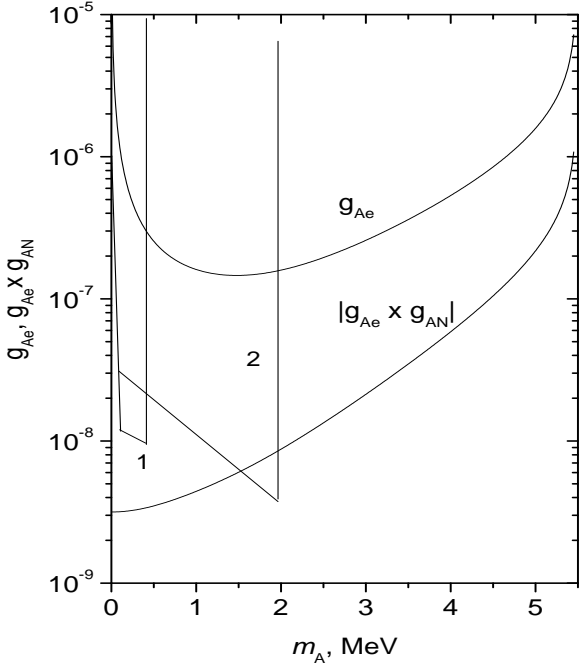}
\caption {Range of excluded $g_{Ae}$ and $|g_{Ae}\times g^3_{AN}|$ values (above the curves), depending on $m_A$, as compared to
the results of the (1) Borexino \cite{Bel08} and (2) Texono \cite{Cha07} experiments. The allowed $g_Ae$ and $|g_{Ae}\times
g^3_{AN}|$ values lie below the corresponding curves. } \label{fig5}
\end{figure}

Within the hadronic axion model, $g_{AN}$  and $m_A$ quantities are related by expression (\ref{gan3}), which can be used to obtain a
constraint on the $g_{Ae}$ constant, depending on the axion mass (Fig. 5). For $m_A$ = 1 MeV, this constraint corresponds to $g_{Ae}\leq
1.8 \times 10^{-7}$. Figure 5 also shows the constraints on the constant $g_{Ae}$ that were obtained in the Borexino experiment for 478-keV
solar axions \cite{Bel08} and in the Texono reactor experiment \cite{Cha07}, where an attempt was made to observe the Compton conversion of
2.2-MeV axions produced in the $n + p \rightarrow d + A$ reaction.

The experiment's sensitivity to the constant $g_{Ae}$ depends on the target mass $M$, specific background level $B$, detector resolution
$\sigma$, and measurement time $T$:
\begin{equation}
g_{Ae}/\delta(g_{Ae})=(MT/\sigma B)^{1/4}.
\end{equation}
The record level of sensitivity $g_{Ae}\approx 10^{-8}$ can be achieved with the detector mass enlarged by an order of magnitude and the
background level reduced by two orders of magnitude (the latter can be done by placing the setup in an underground laboratory).

\section{CONCLUSIONS}
%%%%%%%%%%%%%%%%%%%%%%%%%%%%%%%%%%%%%%%%%%%%%%%%%%%%%%%%%%%%%%%%%%%%%%%%%%%%%%%%%%%%%%%%%%%%%%%
A search for the reaction of the axioelectric absorption of 5.5 MeV axions produced in the $p + d \rightarrow {^3{\rm{He}}} +
\gamma$ (5.49 MeV) reaction was conducted using two BGO detectors with a total mass of 1.2 kg, located in a low-background setup
equipped with passive and active shieldings. As a result, the constraints on the axion-electron coupling constant $g_{Ae}\leq
(1.8 - 9.0) \times 10^{-7}$ for axions with masses $0.1 < m_A < 1$ MeV were obtained. For axions with masses $m_A > 2m_e$, we
calculated the solar positron flux due to the decay of axions into electrons and positrons:$ A \rightarrow e^- + e^+$. The
measured value of the interplanetary positron flux allowed us to establish a new constraint on the axion-electron coupling
constant: $g_{Ae}\leq (1-5) \times 10^{-17}$ for axions with masses in the range of ($1.2-5.4$) MeV.

\section{ACKNOWLEDGMENTS}
%%%%%%%%%%%%%%%%%%%%%%%%%%%%%%%%%%%%%%%%%%%%%%%%%%%%%%%%%%%%%%%%%%%%%%%%%%%%%%%%%%%%%%%%%%%%%%%
We are grateful to E.A. Kolomenskii and A.N. Pirozhkov for supplying the BGO detectors.


\begin{thebibliography}{21}

 \bibitem{Wei78} S.~Weinberg, Phys.~Rev.~Lett. 40, 223 (1978).

 \bibitem{Wil78} F.~Wilczek, Phys.~Rev.~Lett. 40, 279 (1978).

 \bibitem{Pec77} Peccei, R.D. and Quinn, H.R., Phys.~Rev.~Lett.38 1440 (1977).

 \bibitem{PDG08} C.~Amsler et al., (Particle Data Group) Phys.~Lett. B667, 1 (2008).

 \bibitem{Kim79} J.E.~Kim, Phys. Rev. Lett. 43, 103 (1979).

 \bibitem{Shi80} M.A.~Shifman, A.I.~Vainstein, and V.I.~Zakharov, Nucl.~Phys. B 166, 493 (1980).

 \bibitem{Zhi80} A.R.~Zhitnitskii, Yad. Fiz. 31, 497 (1980) [Sov. J. Nucl. Phys. 31, 260 (1980)].

 \bibitem{Din81} M.~Dine, F.~Fischler, and M.~Srednicki, Phys. Lett. B 104B, 199 (1981).

\bibitem{Kap85} D.B.~Kaplan, Nucl.~Phys. B260, 215 (1985).

\bibitem{Ber01} Z.~Berezhiani, et al., Phys.~ Lett. B500, 286 (2001).

\bibitem{Hal04} L.J.~Hall and T.~Watari, Phys.~Rev. D70, 115001 (2004).

\bibitem{Ser09} A.M.~Serenelli, arXiv:0910.3690 (2009)


\bibitem{Der05} A.V.~Derbin, et al., JETP Lett. 81, 365 (2005).

\bibitem{Der07} A.V.~Derbin, et al., JETP Lett. 85, 12 (2007).

\bibitem{Der09A} A.V.~Derbin, et al., Eur.~Phys.~J. C62, 755 (2009). arXiv: 0906.0256

\bibitem{Der07A} A.V.~Derbin, et al., Bull.~Rus.~Acad.~Sci.~Phys. 71, 832 (2007).

\bibitem{Der09} A.V.~Derbin, et al., Phys.~Lett. B678, 181 (2009). arXiv: 0904.3443


\bibitem{Hax91} W.C.~Haxton and K.Y.~Lee, Phys. Rev. Lett. 66, 2557 (1991).

\bibitem{Don78} T.W.~Donnelly, et al., Phys.~Rev. D18, 1607 (1978).

\bibitem{Avi88} F.T.~Avignone III, et al., Phys.~Rev. D 37, 618 (1988).

\bibitem{Sre85} M.~Srednicki, Nucl.~Phys., B260, 689 (1985).

\bibitem{Sch97} G.J.~Schmid et al., Phys.~Rev. C56, 2565 (1997).

\bibitem{Zhi79} A.R.~Zhitnitskii and Yu.I.~Skovpen', Yad. Fiz., 29b, 995 (1979).

\bibitem{Raf82} G.~Raffelt, L.~Stodolsky, Phys.~Lett. B119, 323 (1982).

\bibitem{Fel98} G.J.~Feldman  and R.~Cousins, Phys. Rev. D57, 3873 (1998).

\bibitem{Bel08} G.~Bellini et al., (Borexino coll.) EPJ, C54, 61 (2008).

\bibitem{Cha07} H.M.~Chang et al., (Texono Coll.) Phys.~ Rev. D75, 052004 (2007).


\end{thebibliography}
\end{document}